\documentclass[twocolumn, switch]{article}
\usepackage{hyperref}	% Color links to references, figures, etc.
\hypersetup{colorlinks = true,
            linkcolor = purple,
            urlcolor  = blue,
            citecolor = cyan,
            anchorcolor = black}
\usepackage[backend=biber,doi=false,isbn=false,url=false,eprint=false,style=nature]{biblatex}
\usepackage{preprint}
\usepackage[utf8]{inputenc}	% allow utf-8 input
\usepackage[T1]{fontenc}	% use 8-bit T1 fonts
\usepackage{booktabs} 		% professional-quality tables
\usepackage{nicefrac}		% compact symbols for 1/2, etc.
\usepackage{microtype}		% microtypography
\usepackage{lineno}		% Line numbers
\usepackage{lipsum}		%  Filler text
\usepackage{multicol}
\usepackage{graphicx}
\usepackage{float}			% Allows for figures within multicol
\usepackage{stfloats}
\usepackage[labelsep=period]{caption}
\usepackage{lineno}
\usepackage{pdfpages}
\usepackage{xcolor}		% colors for hyperlinks
\usepackage{newfloat}
\DeclareFloatingEnvironment[name={Supplementary Figure}]{suppfigure}
\usepackage{sidecap}
\sidecaptionvpos{figure}{c}
\usepackage{titlesec}
\titlespacing\section{0pt}{12pt plus 3pt minus 3pt}{1pt plus 1pt minus 1pt}
\titlespacing\subsection{0pt}{10pt plus 3pt minus 3pt}{1pt plus 1pt minus 1pt}
\titlespacing\subsubsection{0pt}{8pt plus 3pt minus 3pt}{1pt plus 1pt minus 1pt}
\usepackage{subcaption}
\titleformat*{\section}{\large\bfseries}
\titleformat*{\subsection}{\normalsize\bfseries}

\title{Deep learning the atmospheric boundary layer height}

\usepackage{xwatermark}
\newwatermark[firstpage,color=gray!90,angle=0,scale=0.28, xpos=0in,ypos=-5in]{*correspondence: \texttt{john.reina@correounivalle.edu.co}}

\usepackage{authblk}

\author[1,2]
{David R. Vivas}
\author[1]{Estiven Sánchez}
\author[1,2 \thanks{\tt{john.reina@correounivalle.edu.co}}]{John H. Reina}
\affil[1]{Centre for Bioinformatics and Photonics (CIBioFi), Universidad del Valle, Edificio E20 No.~1069, 760032 Cali, Colombia}
\affil[2]{Departamento de F\'isica, Universidad del Valle, 760032 Cali, Colombia}

\begin{document}
%\floatsetup[figure]{style=plain,subcapbesideposition=top}

% \linenumbers

\twocolumn[ % Method A for two-column formatting
 \begin{@twocolumnfalse} % Method A for two-column formatting
  
\maketitle

%%%%%%%%%%%%%%%
\begin{abstract}
{\textbf{A question of global concern regarding the sustainable future of humankind stems from the effect due to aerosols on the  global climate. The quantification of atmospheric aerosols and their relationship to climatic impacts are key to understanding the dynamics of climate forcing and to improve our knowledge about climate change. 
Due to its response to precipitation, temperature, topography and human activity, one of the most dynamical atmospheric regions is the atmospheric boundary layer (ABL): ABL aerosols have a sizable impact on the evolution of the radiative forcing of climate change, human health, food security, and, ultimately, on the local and global economy.
The identification of ABL pattern behaviour requires constant monitoring and the application of instrumental and computational methods for its detection and analysis. Here, we show a new method for the retrieval of ABL top arising from 
light detection and ranging (LiDAR) signals, by training a convolutional neural network in a supervised manner; forcing it to learn how to retrieve such a dynamical parameter 
on real, non-ideal conditions and in a fully automated, unsupervised way. 
%The neural-network results here reported have been obtained from data acquired from in-situ measurements performed at the LALINET network LiDAR-CIBioFi station (Cali, Colombia). 
Our findings pave the way for a full integration of LiDAR elastic, inelastic, and depolarisation signal processing, and provide a novel approach for real-time quantitative sensing of aerosols.}}
\end{abstract}
%%%%%%%%%%%%%%%%%

  \end{@twocolumnfalse} % Method A for two-column formatting
] % Method A for two-column formatting

%%%%%%%%%%%%%%%  Main text   %%%%%%%%%%%%%%%
% \linenumbers

%%%%%%%%%%%%%%%%%%%%%%
%\section*{Introduction}
%%%%%%%%%%%%%%%%%%%%%%

Atmospheric particulate matter (PM) stands for a mixture of microscopic solid particles and liquid droplets suspended in the air, which can be of varied chemical composition and size distribution~\supercite{putaud_european_2010}. 
These include coarse and fine particles such as PM$_{10}$, PM$_{2.5}$, including nitrates, sulfites, organic carbon and black carbon~\supercite{isaksen_atmospheric_2009,ramanathan_global_2008,andrews_climatology_2011,sida}. 
Also referred to as atmospheric aerosols, they are a key indicator of air pollution, and their heterogeneity arises from the numerous sources and varying formation mechanisms, as they
can be either directly emitted to the atmosphere or produced in the atmosphere from precursor  gases~\supercite{andreae_strong_2005,larson_decadal_2020,noauthor_line_2020,amigo_when_2020,huebert_overview_2003,kanakidou_organic_2005,knutti_equilibrium_2008}.
Continuous high concentration levels of atmospheric pollution pose many crucial challenges and important adversarial effects, ranging from impact on global public health, mortality and morbidity due to increased cases of cardiovascular and cardio-respiratory diseases linked to long-term exposure to fine particles~\supercite{hoek_long-term_2013,vandyck_air_2018,lelieveld_contribution_2015}, through to direct implications on climatic changes~{\supercite{noauthor_line_2020,andreae_strong_2005,amigo_when_2020,phillips_race_2020,larson_decadal_2020,noauthor_climate_nodate,isaksen_atmospheric_2009,huebert_overview_2003,ramanathan_global_2008,kanakidou_organic_2005,andrews_climatology_2011,knutti_equilibrium_2008}}, 
and agriculture~\supercite{vandyck_air_2018,lipper_climate-smart_2014,vermeulen_addressing_2013,wheeler_climate_2013,schmidhuber_global_2007}. It has become a latent need to accurately monitor the atmospheric variables that allow for the prediction of air pollution behaviour, in order to issue early alarms for the protection of the population. One of the variables of greatest interest is the dynamical height of the atmospheric boundary layer (ABL), the lowest layer of the troposphere that is directly influenced by the Earth surface by means of both natural and anthropogenic emissions~\supercite{Monteith1980}.

The ABL top (ABLT) is defined as the midpoint of the sharp transition zone between the ABL and the free troposphere~\supercite{toledo_estimation_2017}. In an ideal LiDAR signal, it is identifiable as the midpoint of the first sharp reduce of intensity, the so-called entrainment zone. In a real LiDAR signal, several intensity peaks can be detected before or after the actual ABLT, thus erroneously estimating its location.
The fully automated and unsupervised detection of the ABLT has been a challenging topic for the LiDAR community during more than two decades, with the development of numerous contributions such as the gradient~\supercite{comeron_wavelet_2013} and second derivative~\supercite{menut_urban_1999} methods, the wavelet Covariance transform (WCT) method~\supercite{gamage_detection_1993,brooks_finding_2003,mao_determination_2013,baars_continuous_2008}, several fitting methods~\supercite{steyn_detection_1999,lange_atmospheric_2014,cespedes_first_2018}, and some approaches based on the statistical analysis of the LiDAR signal, {over the time evolution of successive signals}~\supercite{melfi_lidar_1985,toledo_cluster_2013}. 
Of the aforementioned, the gradient, second derivative, and WCT have achieved big spread and acceptance~\supercite{toledo_estimation_2017, gamage_detection_1993, brooks_finding_2003, mao_determination_2013, baars_continuous_2008, comeron_wavelet_2013}; probably due to their simplicity and ease of implementation. However, these methods are heavily limited by the weather conditions that determine the shape of the signal, usually failing to retrieve plausible predictions under the presence of clouds, heavy noise and  layer overlapping phenomena. In order to avoid these error sources, a search threshold can be defined for each specific case study, thus sacrificing automaticity in the process. 

Fitting methods attempt to circumvent these limitations via the automatic fitting of the experimental signal to an ideal one, from which an ABLT value~\supercite{steyn_detection_1999} or a search threshold~\supercite{cespedes_first_2018} can be extracted. Nonetheless, these methods are still limited by the shape of the signal, given that under non-ideal conditions, such shape can be substantially different from the ideal one. This translates into the retrieval of an invalid fitting and a consequent inaccurate ABLT value. %, independently from the performance of the optimizer.

The above-mentioned methods assume that the aerosol concentration inside the ABL is significantly higher than the one in the free troposphere~\supercite{toledo_estimation_2017}. These methods process each signal independently, and are computationally inexpensive (assuming an adequate implementation), allowing the possibility of real-time detection. Methods based on statistical analysis usually rely on information of the signal's time evolution over a set of measurements, so they are outside the scope of this work.

In this article, we present a new method for real-time, fully automated and unsupervised ABLT detection from atmospheric LiDAR signals. Our method uses a deep learning model, specifically a convolutional neural network (NN), trained to detect the atmospheric boundary layer top on real, non-ideal conditions. 
The results here reported have been obtained from data acquired from in-situ  measurements performed at the LALINET network LiDAR-CIBioFi station (Cali, Colombia), which has been operational since 2018, and constitutes a regional strategy that contributes to the analysis and prediction of climate, weather and air quality~\supercite{cespedes_first_2018}.

\begin{figure*}[!ht]
\centering
\includegraphics[width=0.9\textwidth]{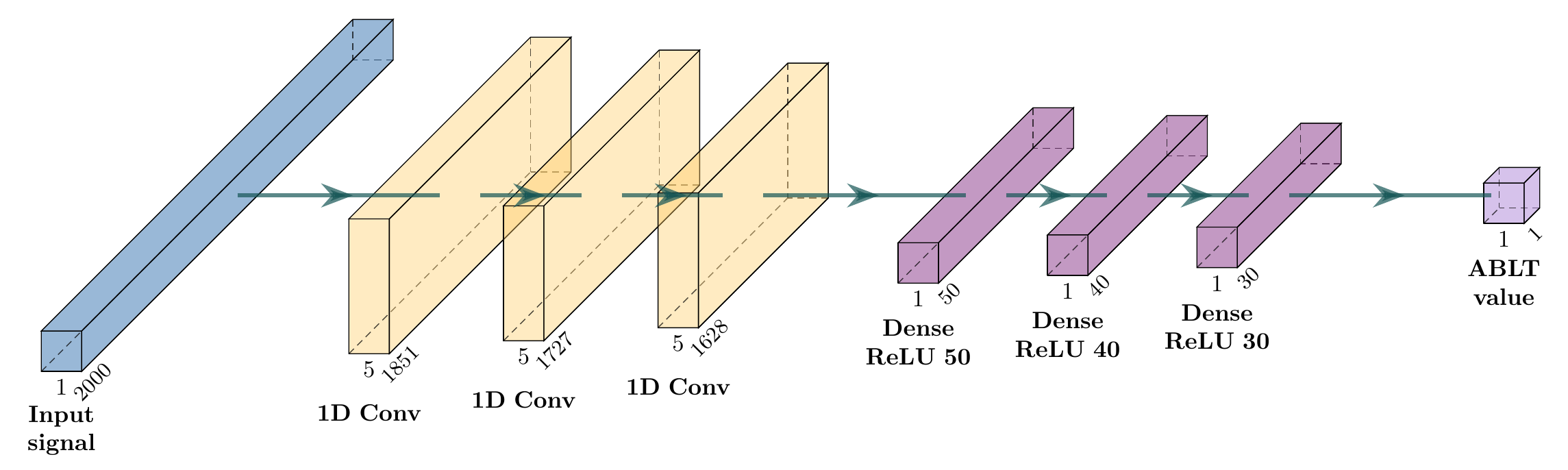}
\caption{\textbf{Neural Network Architecture.} The input signal is fed to three convolutional layers, each one with five kernels that decompose the signal into low-levels features. This new representation of the signal is then fed to a three-layer multi-layer perceptron (MLP), which defines a non-linear transformation that finally returns the detected ABLT. Initially only dense architectures (MLPs) were considered, but their inclination to overfitting lead to poor performance on validation and test data. Progressively increasing the number of convolutional layers at the start of the model significantly increased its generalisation capabilities.}% \sugb{Propagation time through the NN is of the order of milliseconds on a CPU and even faster on a GPU, thus allowing the possibility of real-time detection.}}
\label{fig:architecture}
\end{figure*}

\section*{Results}

\subsection*{Extracting  the atmospheric boundary layer height from LiDAR retrievals: A deep learning approach.} 
Given a range-corrected LiDAR signal (RCS), an ABLT value is obtained by forward-propagating the signal over the layers of a convolutional network~\supercite{lecun_gradient-based_1998}. The architecture of the implemented neural network is illustrated (by means of  PlotNeuralNet~\supercite{iqbal_2020}) in figure~\ref{fig:architecture}. 
The dataset used for the training and tuning of the model is conformed by 15000 signals labeled via WCT, {as explained in Supplementary Material}, with a custom search threshold for each individual one; this was done in order to ensure the quality of the labels and, consequently, the predictions of the neural network. WCT was chosen as the labeling method both for its ease of implementation, and well-known robustness and performance~\supercite{brooks_finding_2003,mao_determination_2013,baars_continuous_2008,bravo-aranda_new_2017}. Assuming the correct training of the model, it is then expected that it \emph{replicates the predictions of a signal-by-signal fine-tuned WCT}, but \emph{in a completely automated and non-supervised manner}.

The convolutional neural network here proposed for ABLT detection is compared to WCT in a supervised variant (custom search threshold for all time evolution) and  unsupervised WCT (full signal as input during all time evolution). WCT has been chosen as the baseline
as for the past two decades it has been used to measure the ABLT in numerous case studies~\supercite{brooks_finding_2003,baars_continuous_2008,bravo-aranda_new_2017} with one of these being Wuhan~\supercite{mao_determination_2013}, a region of growing environmental interest during the 2020 
SARS-CoV-2 virus
%(COVID-19) 
pandemic outbreak~\supercite{covid19}.
%%%%%%%%%%%%%%%%%%%%
\begin{figure*}[b]
\centering
\begin{subfigure}{1.0\linewidth}
\caption{}
\includegraphics[width=1.0\textwidth]{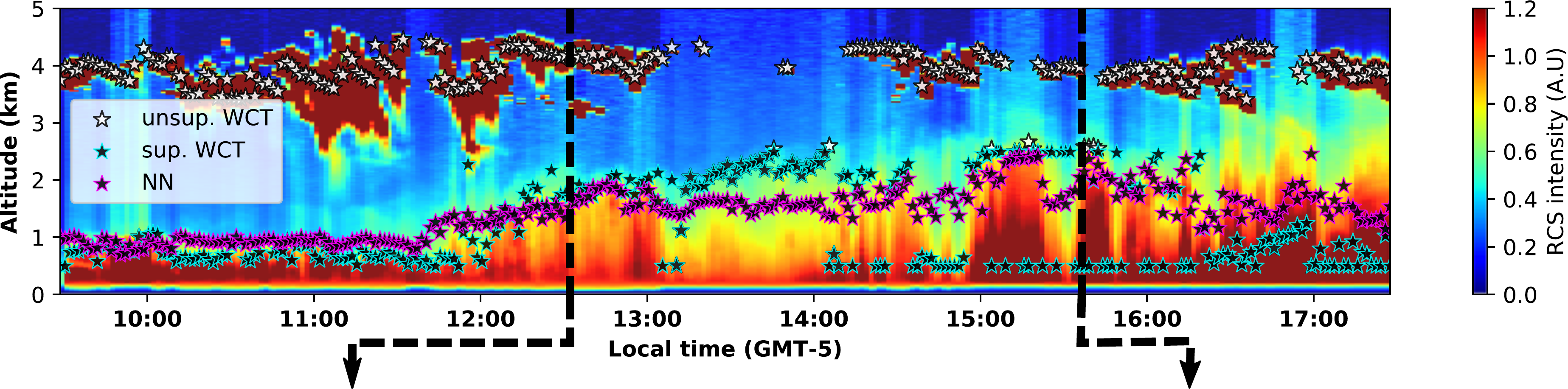}
\label{fig:temporal1}
\end{subfigure}
\par
\begin{multicols}{2}
    \begin{subfigure}{0.9\linewidth}
   \vspace{-0.5cm} 
   \hspace{-1.5cm}
  \caption{}  
  \vspace{-0.4cm}
    \hspace{0.0cm}
\includegraphics[width=0.9\linewidth]{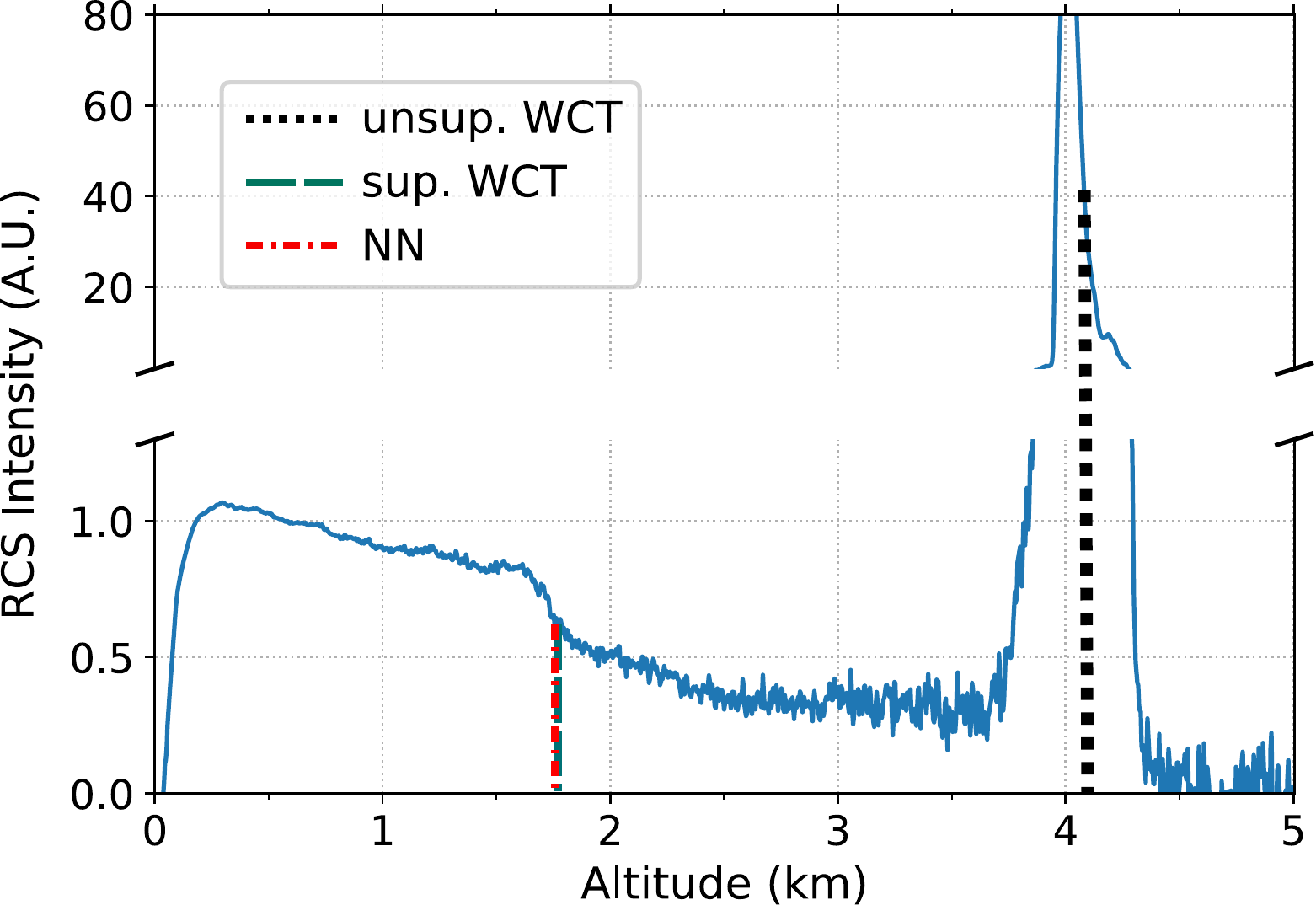}
\label{fig:profile1}
\end{subfigure}
\par
\hspace{-1.2cm}
\begin{subfigure}{0.9\linewidth}
   \vspace{-0.5cm} 
   \hspace{-1.5cm}
\caption{}
  \vspace{-0.4cm}
    \hspace{-0.2cm}
\includegraphics[width=0.9\linewidth]{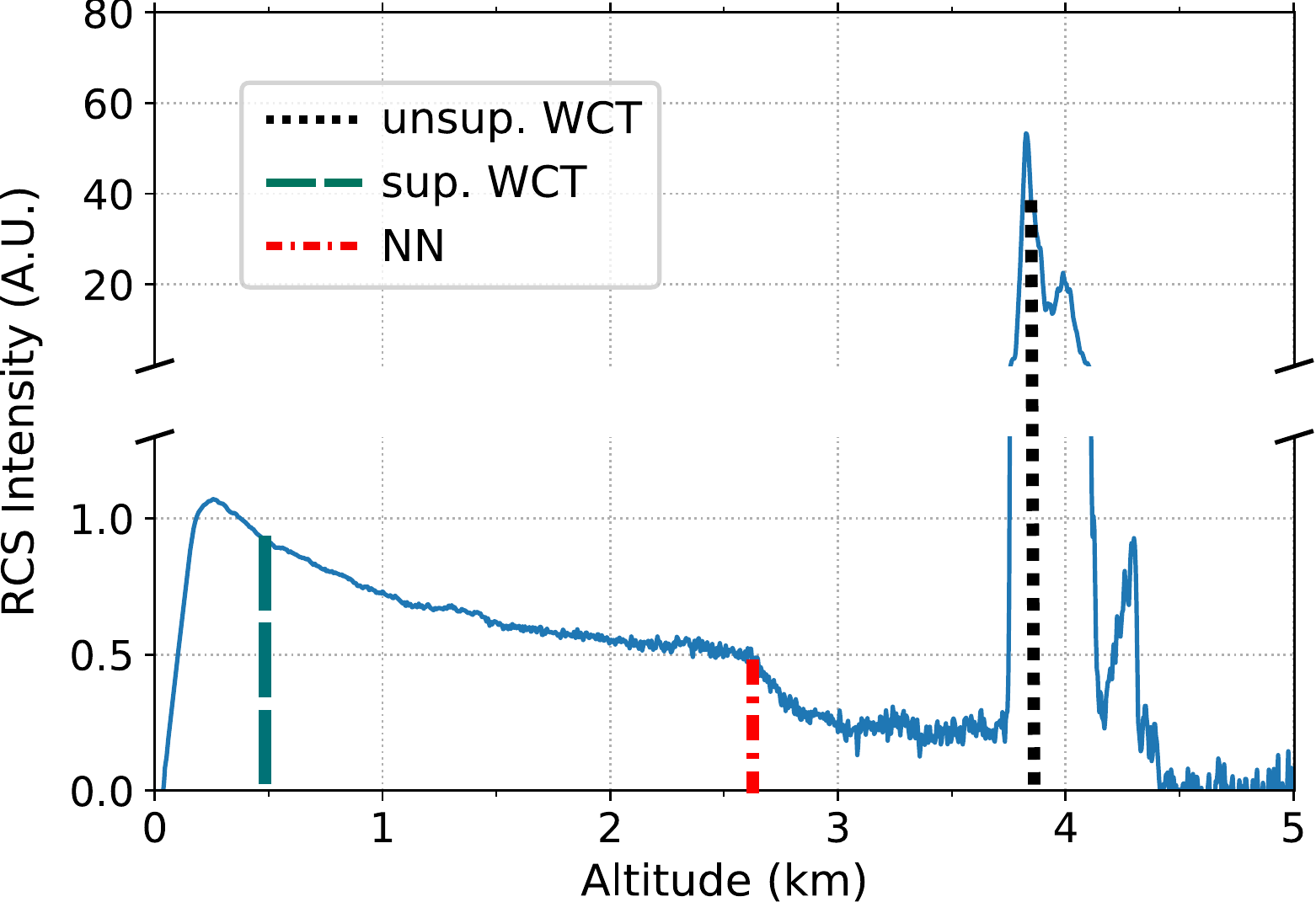}
\label{fig:profile2}
\end{subfigure}
\end{multicols}
    \caption
{\textbf{LiDAR retrievals for \textit{august 14, 2019}}. {\textbf{a}},  Temporal evolution of the ABL, and {\textbf{b}}, {\textbf{c},} Two selected single LiDAR profiles. They point out  different scenarios treated with our method: the measurements exhibit {\textbf{b},} Profile 1 (12:30 h), a well-mixed layer where NN and sup.~WCT values are very similar, and {\textbf{c}}, Profile 2 (15:40 h), unusual atmospheric conditions where the NN estimation differs from both sup. and unsup.~WCT values; the latter profile exhibits  a likely turbulence, with a height about 2.6 km for the NN prediction. The intensity of the signals is given in arbitrary units (a.u.).}
    \label{fig:aug14profiles}
\end{figure*}

%%%%%%%%%%%%%%%%%%%%%%%%%%
\begin{figure*}[!ht]
    \begin{subfigure}{1.0\textwidth}
\caption{}
\includegraphics[width=1.0\textwidth]{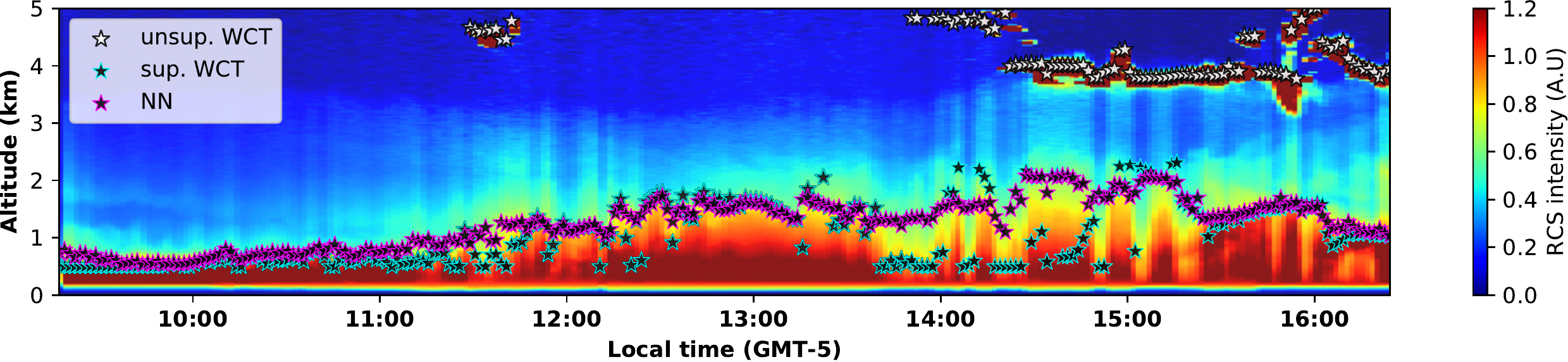}
\label{fig:aug152019}
\end{subfigure}
    \begin{subfigure}{1.0\textwidth}
\caption{}
\includegraphics[width=1.0\textwidth]{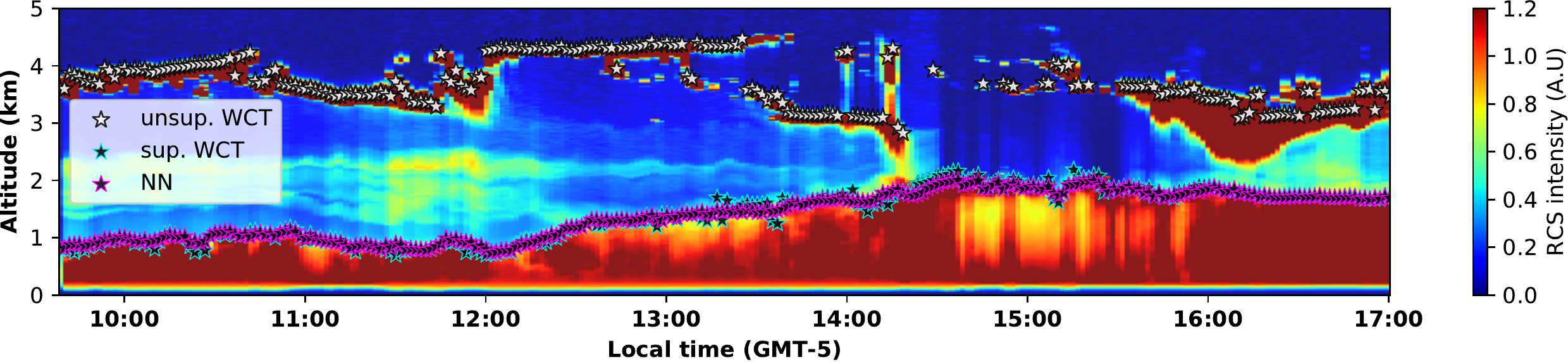}
\label{fig:aug162019}
\end{subfigure}
    \begin{subfigure}{1.0\textwidth}
\caption{}
\includegraphics[width=1.0\textwidth]{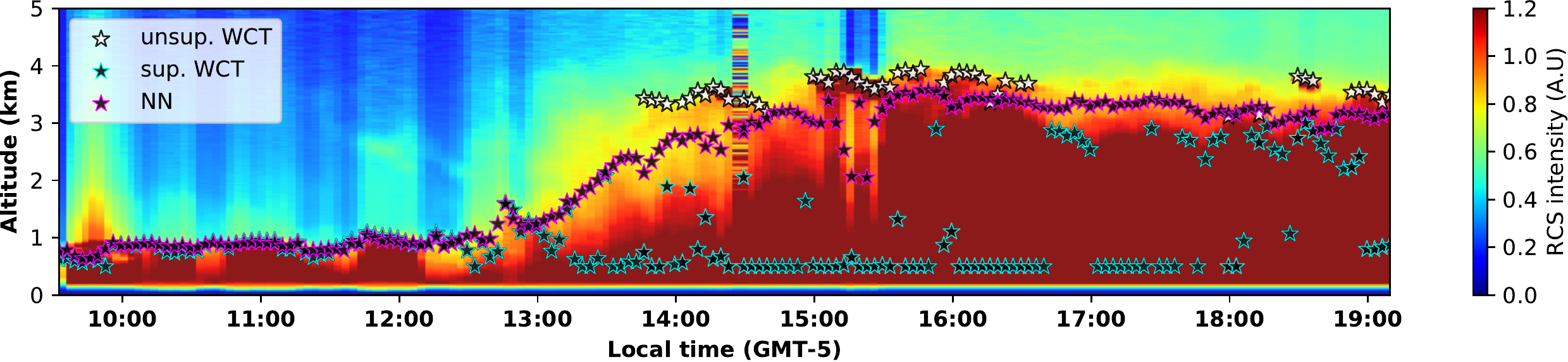}
\label{fig:aug212019}
\end{subfigure}
\caption
{\textbf{Case study, 
\textit{august 2019} (Amazon rain forest fires).}
{\textbf{a}}, August 15, {\textbf{b}}, August 16, {\textbf{c}}, August 21, 2019.
These measurements exhibit unusual atmospheric conditions, with ABLT values of up to four kilometers in height and dense cloud layers close to the ABL. Here the generalisation capabilities of the NN are proven, as the neural network provides ABLT predictions of continuous time evolution and low variance (figure \ref{fig:hist}), and within the region associated to the entrainment zone.}
\label{fig:aug2019}
\end{figure*}

Over the past six months,
fires of unprecedented impact on people wellbeing, biodiversity, wildlife, and infrastructure have dramatically affected the australian continent (january 2020)~\supercite{noauthor_line_2020,phillips_race_2020} and the Amazon rainforest (august 2019)~\supercite{amigo_when_2020, andrade_alarming_2019}. Their possible connection to climate changes~\supercite{phillips_race_2020,staal_forest-rainfall_2018,noauthor_climate_nodate} have raised serious alarms, and a recurrent need for global environmental policy and timely law reinforcement, as well as for more accurate climate dynamics and atmospheric predictions, all of which constitute a crucial challenge for both adaptation and mitigation~\supercite{phillips_race_2020,staal_forest-rainfall_2018,noauthor_climate_nodate}.

A detailed comparison of the results for ABLT prediction on different days (14, 15, 16 and 21) of august 2019, from data obtained at the LiDAR-CIBioFi station, is  presented in figures \ref{fig:aug14profiles} and \ref{fig:aug2019}. During such period, strong aerosol transport from the
Amazon rain forest wildfires was reported by NASA's Atmospheric Infrared Sounder (AIRS) instrument aboard the Aqua satellite~\supercite{NASA_2019}, showing the aerosols propagation from the source (brazilian northwest Amazon region) through to the north of South America, during 8-22 august 2019, thus providing a challenging test bench for the above-described methods. 
The size of NASA's reported fires 
 was so large that they could be spotted from space, and spread over several large Amazon states in northwest Brazil\supercite{andrade_alarming_2019,amigo_when_2020,staal_forest-rainfall_2018}.
The results are depicted in graphs of temporal evolution (see figures \ref{fig:aug14profiles} and \ref{fig:aug2019}) that exhibit very different LiDAR results throughout the day.  The dynamics are presented in 2D graphs, with the signal intensity plotted in a colour scale; in figure \ref{fig:temporal1}, each dashed vertical line corresponds to a single LiDAR measurement profile, which in turn are shown in figures~\ref{fig:profile1}, and~\ref{fig:profile2}.

On August 14 (figure \ref{fig:aug14profiles}),  clouds around 4 km height, with some formations around 2 km were detected, with the latter being very close to the height of the boundary layer, thus posing a challenge for accurate ABLT detection. In addition, some cases of turbulence were detected after 14 h. The first single LiDAR measure profile (figure~\ref{fig:profile1}), taken at about 12:30 h, exhibits a stable and well-mixed layer, making it easy to discriminate between ABL and free troposphere, thus allowing a straightforward evaluation of the predictions: the NN gives very similar results to the supervised WCT (about 1.8 km), while the unsupervised WCT located the ABLT in a cloud formation above 4 km height. In contrast, the second profile (figure~\ref{fig:profile2}), taken at 15:40 h, gives very different results for the supervised (sup.) WCT, the unsupervised (unsup.) WCT, and the NN. The sup.~WCT located the ABLT at 500 m, below the expected result. The unsup.~WCT placed the result at around 4 km, in a cloud formation pattern, while the NN located the ABLT about 2.6 km, following its actual behaviour. The temporal evolution of the LiDAR measurement profiles of figure~\ref{fig:aug14profiles} makes it clear that the unsup.~WCT detection locates the position of the boundary layer at cloud formation height, erroneously placing the ABLT  in most cases. The sup.~WCT shows ABLT detection problems, severely underestimating ABLT, for cases of proximity to clouds (see e.g., around 12 h), and in cases of greater turbulence (e.g., after 14 h). Despite the drawbacks presented by sup.~WCT and unsup.~WCT, figure~\ref{fig:aug14profiles} clearly shows that our convolutional NN estimation of ABLT is resilient to turbulence and proximity of clouds, and correctly follows its evolution.

\subsection*{Aerosol transport and  biomass-burning emissions identification: The case of 2019 Brazilian rainforest wildfires.}
Following our analysis, we next introduce the results for the case study of the Amazon rain forest fires (august 2019), as plotted in figure~\ref{fig:aug2019},  for august 15, 16, and 21. 
August 15 (figure \ref{fig:aug152019}) is a day of particular interest, since we  detected ABLT levels around 0.5 km in the morning, well below the 1 km height--approximately the average value in the study region. This decrease in ABLT corresponds to an increase in the PM concentration, possibly due to a temperature decrease in the troposphere. This was corroborated by the local authority for air quality monitoring  (DAGMA), and alerts due to local high PM concentration
%during this period 
were issued on this date~\supercite{dagma2019}.  Similar concerns already pointed out for the ABLT detection methods were encountered in this case, for both supervised and unsupervised WCT. The supervised WCT, for example, registers ABLT values close to 400 m at about 14:00 h, which would imply a PM concentration higher than that reported during the morning, an incorrect prediction following the measurement results. This points out how critical is the requirement of a correct ABLT prediction in assessing the need for communicating early alarms due to high PM concentration, thus avoiding false positives.

Figure~\ref{fig:aug162019} (August 16)  exhibits a stable and well-mixed layer, which allows for a straightforward discrimination between ABL and free troposphere; in this case,  the supervised WCT provides similar predictions to the NN. Unsupervised WCTs, as expected, are located in a layer of cloud formation around 4 km.
%%%%%%%%%%%%%%%%%%%%%
\begin{figure}[!ht]
\includegraphics[width=1\columnwidth]{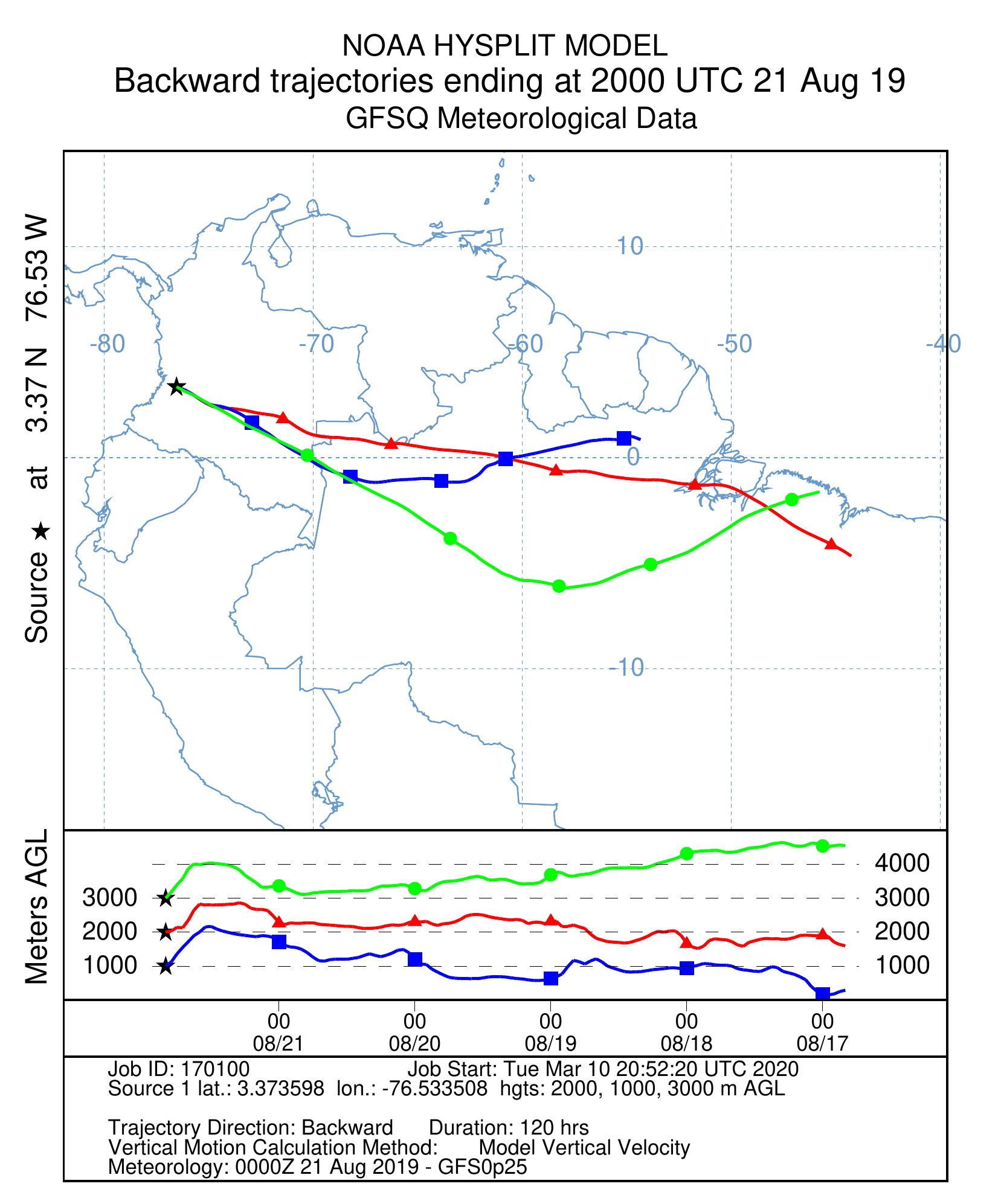}
\caption
{\textbf{NOAA HYSPLIT Backward trajectories run in ensemble mode.}
Five days records and trajectories finishing at 1 km, 2 km and 3 km AGL at 20:00 UTC, on 21/08/2019. The simulation  shows the wind transport from the brazilian northwest amazon rainforest to the place marked with a star, where in-situ measurements were taken at the LiDAR station in Cali (Colombia).}
\label{fig:hysplit}
\end{figure}

On August 21 (figure \ref{fig:aug212019}), an unusual local scenario was detected: ABLT levels about 4 km, well above the  1 km average, with small variations in the PM concentration during the day. A sharp  ABLT  increase of this kind is expected to be followed by a decrease in the PM concentration: if the latter  is not observed, it is most likely that aerosol transport from a different source, say another region, may be occurring. This hypothesis is demonstrated in this case by analysing the wind and aerosol transport by means of the HYSPLIT backward trajectories model (figure~\ref{fig:hysplit}), from which we were able to trace, on this date,  the burning of biomass due to the 2019 Amazon rain forest fire~\supercite{NASA_2019}. In addition, and to corroborate this result, data reported by NASA's Atmospheric Infrared Sounder (AIRS) instrument aboard the Aqua satellite~\supercite{NASA_2019} show that indeed the results presented in figure~\ref{fig:aug212019} coincide with aerosols flow from the brazilian northwest Amazon wildfires~\supercite{NASA_2019} (see Supplementary Material for more details about aerosol transport during 2019 summer Amazon rainforest wildfires).  

% \textbf{Computational methodologies for ABLT detection vs. convolutional NN: Performance and statistical analysis.} 
\subsection*{NN vs WCT: Performance comparison.} 
Predictions for ABLT results arising from the three methods here considered are quite different from each other in the latter scenario (figure~\ref{fig:aug212019}). The unsupervised WCT is only able to locate values near the ABLT during an interrupted time window of about two hours, otherwise its forecasts are above 5 km in upper clouds (not shown). The supervised WCT shows ABLT values well below the actual ones, detecting most of  them below 1 km height during extended time windows, which is an overly underestimation.
In contrast to the previous results, the proposed NN-based method follows the appreciated ABLT evolution contour, during the full time-frame measured (9:30-19:00 h) and without difficulty. The results show that the high expressive power of the neural-network makes it possible to obtain physically plausible ABLT values, even in the presence of unusual atmospheric phenomena that are difficult to predict for traditional methods such as WCT. We stress that the propagation time through the NN is of the order of milliseconds on a CPU, and even faster on a GPU, thus allowing the possibility of real-time detection.
%%%%%%%%%%%%%%%%%%%%%
\begin{figure}[!ht]
\includegraphics[width=1\columnwidth]{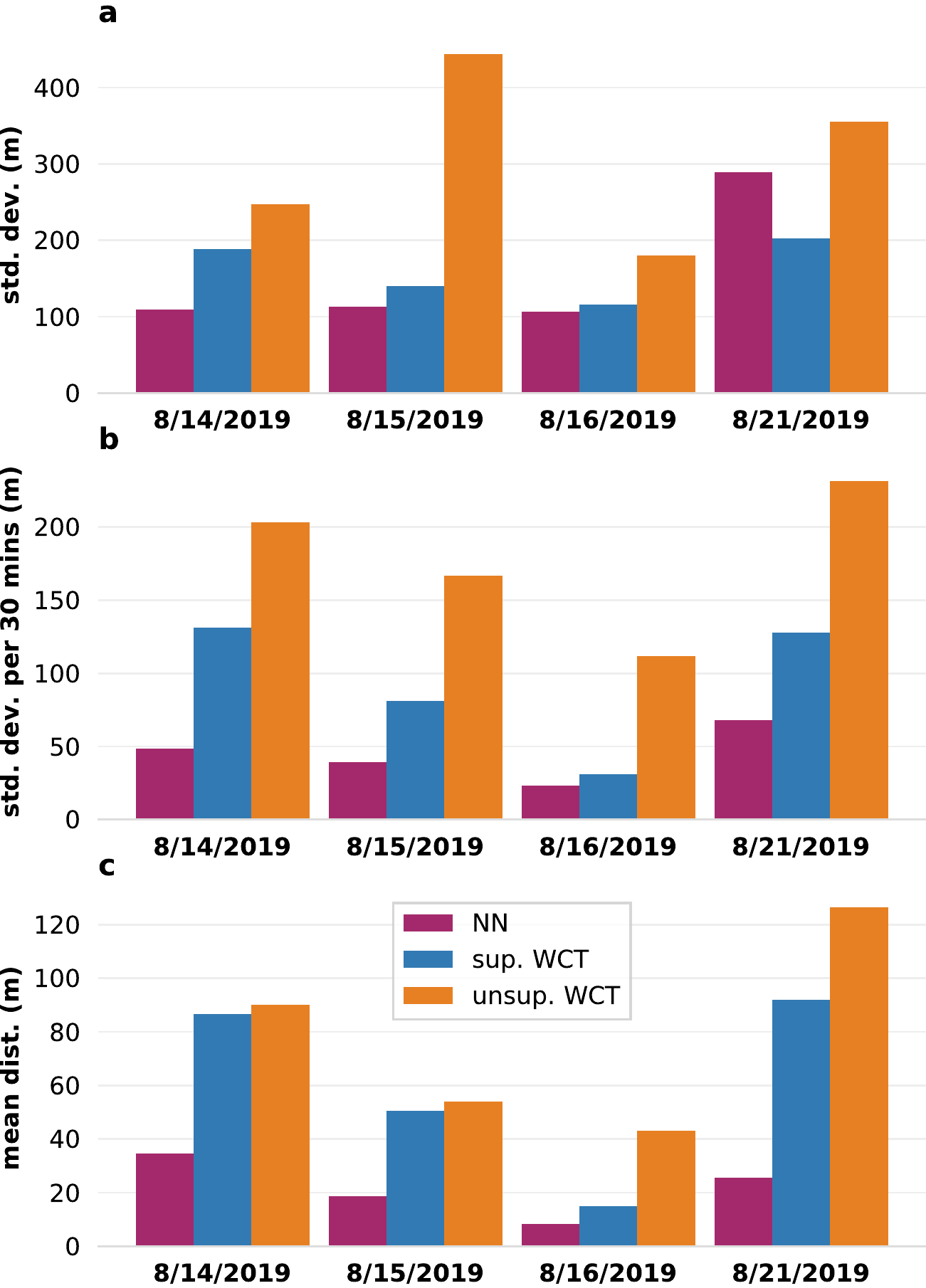}
\caption{\textbf{Statistical results for the performance of sup.~WCT, unsup.~WCT, and the proposed NN model.}
\textbf{a}, Standard deviation of the full ABLT time evolution, \textbf{b}, Mean standard deviation over 30 minute intervals,  \textbf{c}, Mean distance between points for all the case studies.
%(figures~\ref{fig:aug14profiles} and \ref{fig:aug2019}). 
The NN achieves the lowest dispersion and the highest continuity, followed by sup. WCT. 
%As 8/21/2019 case study shows, the full std. dev. is not a reliable continuity metric.
}
\label{fig:hist}
\end{figure}
%%%%%%%%%%%%%%%%%%%%%%%%%

Besides the prediction analysis of every particular case-study, we have performed a measurement of the standard deviation and mean distance between points (see figure \ref{fig:hist}), as a mean to quantify and compare the continuity of the ABLT time evolution provided by each method. 
The statistical results for the performance of supervised~WCT, unsupervised~WCT, and the proposed NN model are shown, for all the case studies, in the histograms, figure \ref{fig:hist}, that plot the following: \textbf{a} Standard deviation of the full ABLT time evolution, \textbf{b} Mean standard deviation over 30 minute intervals, and \textbf{c} Mean distance between points. 

The standard deviation of the full time evolution can be used as a reliable metric of dispersion, but not of continuity. This is illustrated in figure \ref{fig:hist}\textcolor{purple}{a} for August 21, where the full std. dev. of supervised WCT is smaller than the one of the NN, in stark contrast to what is shown in figure~\ref{fig:aug212019}. Given this, the mean distance between each point and its immediate neighbour was considered as a continuity metric, as it provides results that are consistent with those of figures~\ref{fig:aug14profiles} and~\ref{fig:aug2019}. The mean standard deviation over blocks of 30 minutes acts as a middle ground metric between the aforementioned ones, as it quantifies dispersion on a small time window according to the high temporal variability exhibited by the ABL~\supercite{stull}.

The results show that, for all the case studies here treated, the tested NN achieves the highest continuity of time evolution and least dispersion on its ABLT predictions, followed by sup. WCT. This coincides with the dynamical behaviour observed in figures~\ref{fig:aug14profiles} and~\ref{fig:aug2019}.

%%%%%%%%%%%%%%%%%%%%%%%%
\section*{Discussion}
%%%%%%%%%%%%%%%%%%%%%%%%

We conclude by summarising our findings and perspectives. A novel technique for active atmospheric remote sensing, and, in particular, for atmospheric boundary layer estimation has been presented using convolutional neural networks, having the data processed in real time, and without the need for supervision or postprocessing.
A large set of LiDAR data were collected in-situ with instrumentation available at our station, and their corresponding boundary layer values labeled in a controlled manner with WCT. To validate and evaluate the performance of our model, we have analysed different scenarios, under very different atmospheric conditions.  Compared to the WCT approaches, the proposed NN results were more robust and could readily be used as an ABLT estimator. We found that, to quantify ABLT during these experiments, the structure of NN provided better regression effects than those from WCT. Our network architecture easily adapts different ranges of behaviour of the mixing layer, differentiating turbulence and cloud proximity phenomena. 
%%%%%%%%%%%%%%%%%%%%
\begin{figure}[!ht]
\centering
\includegraphics[width=0.75\columnwidth]{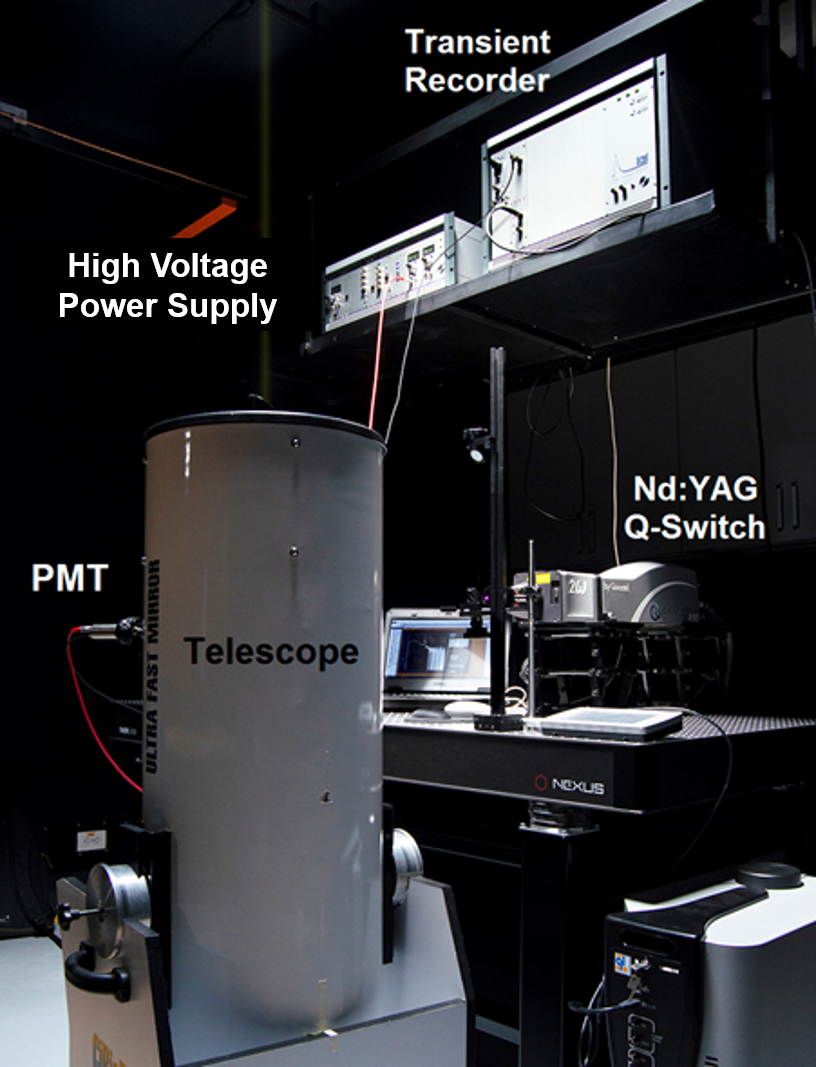}
\caption
{\textbf{Experimental setup.}
Instrumentation at the atmospheric  LiDAR-CIBioFi station (3.37N; 76.53W) in Cali, northwestern South America. The LiDAR configuration operates in elastic mode at 532 nm; the  Nd:YAG pulsed laser source %(Q-smart 450 mJ by Quantel)
features a pulse width of $\leq$ 6 ns at 1064 nm, and a pulse repetition frequency of 10 Hz.}
\label{fig:LiDAR}
\end{figure}

As a test bed for our model, we have considered an example of important complexity: we have reported a several days scenario during August 2019, with well marked different behaviour and particularities. This time frame coincided with the northwest brazilian summer Amazon rain forest wildfires~\supercite{NASA_2019,andrade_alarming_2019,amigo_when_2020,staal_forest-rainfall_2018}. We have shown how the reported NN model allowed us to quantify the influence due to such wildfires on our local aerosol dynamics: despite turbulence in the mixture layer and close presence of clouds, we were able to successfully  identify the ABLT by means of the proposed neural network. 

Our model has been specifically trained  with 532 nm  LiDAR data in elastic configuration, but future work will consider its extension to additional information channels, such as other elastic wavelengths and inelastic scattering, as well as to the depolarisation channel~\supercite{bravo-aranda_new_2017}. As the neural networks excel at learning highly complex correlations between channels of data, the implementation of such extensions would lead to an improved accuracy, since this can be used both as an additional input for the neural network, and as a complementary method for refining the labels fed to the model during the training time.

% {\large{\textbf{Methods}}}
\section*{Methods}

% {\big{\textbf{Experimental setup}}}
\subsection*{Experimental setup}

The experimental setup that provided the LiDAR retrievals used in this work is shown in figure~\ref{fig:LiDAR}. It consists in a LiDAR system located at CIBioFi-Universidad del Valle (3.37N; 76.53W). The LiDAR configuration used operates in elastic mode to receive single-scattered signals at 532 nm of wavelength, with a coaxial static alignment. The emission component comprises a single frequency Nd:YAG pulsed laser (Q-smart 450 mJ, by Quantel) with a pulse width of $\leq$ 6 ns at 1064 nm, and a pulse repetition frequency of 10 Hz. A second harmonic generator (SHG) was implemented in order to double the central frequency resulting in a wavelength of 532 nm with an average energy $\geq$ 220 mJ. The output beam has a diameter of 6.5 mm which is expanded three times ($3\times$) through a Galilean system.

A Newtonian telescope with a focal length of 1.0 m and a primary mirror with a diameter of 0.3 m constitutes the reception component. The telescope with a field view of 1.47 mrad collects the backscattered light by the atmosphere. The ocular is coupled to a diaphragm with variable aperture, then to an interferential filter at 532 nm, and finally to a photomultiplier tube (Hamamatsu, R9880U series) to guarantee the reception of the light elastically scattered with a quantum efficiency of 50\% at 532 nm.

The transient recorder system (Licel, TR20-160) comprises the configuration detection unit. The transient system allows the discrimination of  signals of different altitude through the synchronisation with a periodic signal (the trigger), which generally comes from the Q-switch  control of the laser system, achieving a spatial resolution of 3.75 m in the vertical atmospheric column.

The acquisition protocol consists of routine observations on different days, storing LiDAR signals with at least 2000 pulses per data from 8:00 until 18:00, local time, on a weekly basis, since 2018 up to date.

% {\Big{\textbf{Computational method}}}
\subsection*{Computational method}

Hyperparameter tuning was achieved via importance sampling over progressively smaller hyperparameter grids. Logcosh was chosen  as loss function and Adam~\supercite{kingma_adam:_2017} as optimizer, as they exhibited the fastest convergence for this particular problem. Batch normalisation~\supercite{ioffe_batch_2015} was used between layers for even faster convergence. No L2 regularisation or dropout were used (see Supplementary Material).

Training was performed on 400 epochs with a linear learn rate decay from an initial value 0.1 to a final value of 0.001. 12000 signals were used as a training set, 3000 signals as cross-validation set, and another 2000 unlabeled signals as a qualitative test set, all normalised via standardisation. This computational method was implemented in the TensorFlow library~\supercite{tensorflow} via its Python 3 API. Additional information about the ABLT neural network---ABLT-NN code can be found in Supplementary Code.

\section*{Data availability}

The data reported in this article are available from the corresponding authors upon reasonable request.

%%%%%%%%%%%%%%%%%%%%%%%%%%%%%%%%%%
\normalsize
\printbibliography
%%%%%%%%%%%%%%%%%%%%%%%%%%%%%%%%%%%

\section*{Acknowledgements}

This work was funded by the  Colombian Science, Technology and Innovation Fund-General Royalties System (Fondo CTeI-Sistema General de Regalías) and Gobernación del Valle del Cauca (Grant BPIN 2013000100007), Fundación para la Promoción de la Investigación y la Tecnología  (Grant 201921). We are grateful to the Laboratory for Atmospheric Physics (LAFA) and to the LiDAR station at the Centre for Bioinformatics and Photonics (CIBioFi) for the provided data.

\section*{Author contributions}

D.R.V. wrote the convolutional neural network code, D.R.V. and E.S. ran the model simulations, collected data and prepared the figures. J.H.R. conceived the study and supervised the experiment and simulations. All authors contributed to the analysis and interpretation of the results, and to the writing of the manuscript.

\section*{Additional information}

Supplementary information is available in the online version of the paper. Correspondence and requests for materials should be addressed to J.H.R.

\section*{Competing interests}

The authors declare no competing interests.

\lhead{}
\pagenumbering{gobble}

\clearpage



\section*{Supplementary material}

# Supplementary Material

## Deep Learning the Atmospheric Boundary Layer Height

David R. Vivas, Estiven Sánchez & John H. Reina

This Supplementary Material describes the computational methodologies employed in the main paper and provides additional data and results about the aerosol transport during the 2019 Amazon rain forest wildfires.

## 1 Wavelet covariance transform method

The Wavelet covariance Transform (WCT) method[1,2] is based on the convolution between the LiDAR range-corrected signal $B(z)$ and a Haar wavelet $h$, which is defined as[3]:

$$h\left(\frac{z}{a}\right) = \begin{cases} +1: & -\frac{a}{2} \leq z \leq 0 \\ -1: & 0 \leq z \leq \frac{a}{2} \\ 0: & \text{else,} \end{cases}$$

where the parameter $a$ is known as the amplitude or dilation of the wavelet. The convoluted WCT $(z, a)$ signal is then:

$$\text{WCT}(z, a) = B(z) * h\left(\frac{z}{a}\right). \quad (1)$$

Convolving $B(z)$ with $h\left(\frac{z}{a}\right)$ results in a new signal where each point indicates a degree of similarity between both. Given that $h\left(\frac{z}{a}\right)$ is, essentially, an abrupt gradient, each point $z$ of the convoluted signal will quantify the gradients present on an interval of amplitude $a$ around the point $z$ of the original signal. Thus, the minimum of this convolution represents the point of higher similarity between $B(z)$ and $h\left(\frac{z}{a}\right)$, and its altitude $z_{\text{ABL}}^{\text{wct}}$ is then, assuming that an adequate interval around the entrainment zone was chosen, taken as the atmospheric boundary layer top (ABLT):

$$z_{\text{ABL}}^{\text{wct}} = z_{res}^{lidar} \times \text{argmin}\left(\text{WCT}(z, a)\right), \quad (2)$$

where $z_{lidar}^{res}$ denotes the vertical spatial resolution of the LiDAR setup (3.75 m in our configuration). The dilation $a$ defines the width of each convolution window, so if such dilation matches the entrainment zone width, the convolution region corresponding to it will be maximised relative to gradients of smaller amplitude, such as the ones associated to instrumental noise[2,4], thus providing an ABLT value close to the midpoint of the indicated entrainment zone (assuming an adequate search interval was chosen). Nevertheless, the presence of clouds represents a factor of failure for this method, given that the optical width of these can reflect as gradients higher than the one associated to the entrainment zone, even for dilations close to the entrainment zone width[2].

The method regarded in this work as *supervised* WCT differs only in the fact that the search interval of the $z_{\text{ABL}}^{\text{wct}}$ is limited to a threshold that encompasses the entrainment zone observed for the entire time evolution of each particular case study, in order to approximately avoid clouds or large gradients that can negatively affect the performance of the method, thus requiring supervision. In contrast, *unsupervised* WCT defines a search threshold that goes from the start of the LiDAR signal to an upper bound defined by the observed interpretable portion of the signal, which in our particular experimental setup corresponds to around 6 km of altitude.

It is worth mentioning that the convolution operation involved in this method is treated in its traditional mathematical sense (similar to that of equation (4), but flipping the kernel on its dimensions before the operation), so it is not equivalent to the operation treated as a convolution in section 2.2 of this supplementary material. If the convolution were performed as a cross-correlation operation, the resulting signal would be flipped around the $x$-axis, so argmax should instead be used.

## 2 Artificial neural networks

### 2.1 Multi-layer perceptron

Artificial neural-networks (NNs) are computational systems partially inspired by the biological neuron. The main components of the architecture of a NN are computing units called Artificial Neurons (ANs), which are interconnected in such a way that allows the propagation of information through the structure of the NN. The branch of artificial intelligence that concerns the design, architecture and optimisation of NNs is known as *deep learning*[5,6].

The quintessence of deep learning is the perceptron. Its more general version, the multi-layer perceptron (MLP)[5,7–9], is a sequential model conformed by one or more layers of ANs. An MLP defines a transformation $y = f(X, \theta)$ and is trained to learn the parameters $\theta = \{W, b\}$ that best approximate $f$ to an objective function $f^*$ which is, in general, unknown explicitly, but con-

tained in the inner structure of a set of data[6,9]. MLPs are also called *feedforward neural networks*, because in these, information flows from an *input layer* to the *hidden layers* of ANs, and finally to an *output layer*; there are no feedback connections in which the output of a layer returns to itself as input.

The propagation of information between two sequential layers, say the $(l-1)$-th and the $l$-th, of an MLP is done via the following expression:

$$A^{[l]} = g^l(W^{[l]T} A^{[l-1]} + b^{[l]}), \quad (3)$$

where $A^{[l]}$ is the vector containing the artificial neurons of the $l$-th layer, $W^{[l]}$ is a dense weight matrix which defines the connections between the neurons of both layers, $b^{[l]}$ is a weight vector called bias and $g^{[l]}$ is a non-linear function that is applied element-wise to induce non-linearity in the transformation defined by the MLP. The initial layer $A^{[0]}$ of an MLP corresponds to the input layer $X$.

Given that $W$ is a dense matrix, the computational cost of MLPs scales poorly with the dimensionality of data. To approach this problem, a new NN architecture inspired by the biological visual cortex was conceived, the so-called convolutional neural network.

## 2.2 Convolutional neural networks

In deep learning, *convolution* refers to the operation usually regarded as cross-correlation in mathematics. For the two-dimensional case, this operation involves a matrix $M$ and a filter or kernel $K$, and returns the degree of similarity $C$ between $M$ and $K$, for each of the points of $M$ on that convolution is possible. Mathematically, this operation is denoted by a star "$\star$" and is defined as:

$$C(i,j) = (M \star K)(i,j) = \sum_m \sum_n M(i+m, j+n) K(m,n), \quad (4)$$

where $(i,j)$ are the matrix indices of $M$ and $(m,n)$ those of $K$. For the one-dimensional case of a given signal $S$, the operation is reduced to:

$$S(i) = \sum_m C(i+m) K(m). \quad (5)$$

In deep learning, additional parameters are usually defined on the convolution operation, such that the dimensions $m_1^{conv} \times m_2^{conv}$ of the two-dimensional convolution between an array of dimensions $m_1 \times m_2$ and a filter with dimensions $k_1 \times k_2$ are given by:

$$m_1^{conv} = \frac{m_1 + 2p_1 - k_1}{s_1} + 1 \quad (6)$$

$$m_2^{conv} = \frac{m_2 + 2p_2 - k_2}{s_2} + 1, \quad (7)$$

where the parameters $(s_1, s_2)$ are known as *stride*, and indicate the number of elements to skip between consecutive filter applications. The $(p_1, p_2)$ parameters are known as *padding*, and indicate the number of zeros that will be added around the edges of the input, in order to preserve their original dimensions after convolution.

When a NN layer uses the MLP propagation function, equation (3), it is said to be a *dense layer*. In contrast, when a layer of a NN uses

$$A^{[l]} = g^l(W^{[l]} \star A^{[l-1]} + b^{[l]}) \quad (8)$$

as its propagation function, then it is said to be a *convolutional layer*. Here $W^{[l]}$ is a multidimensional array of filters that are convoluted in batch with the input of the layer. A *Convolutional Neural Network* (CNN, ConvNet or simply NN) is usually defined as a neural network with at least one convolutional layer[6,10,11], although MLPs are sometimes regarded as CNNs.

Unlike dense layers, convolutional layers operate by applying the same set of weights (filters) over each region of the input, and are said to be *spatially regularized*. It is the application of multiple low-dimensional filters that allows a convolutional layer to abstract features of a high-dimensional input (such as high-resolution images) at a low computational cost[6,10,11].

## 2.3 Training the neural network

The parameters of a NN architecture that can be only varied by the programmer (such as the number of hidden layers or the number of neurons per layer) are regarded as *hyperparameters*. The remaining parameters that *can* be varied by the machine (such as the weights and biases) are adjusted via a training process performed by means of an algorithm called *backpropagation*[9].

Given that backpropagation is a delicate and challenging algorithm to implement, especially for deeper architectures, training of neural networks is usually performed via linear algebra and symbolic derivation libraries such as Tensorflow[12] that automatically implement and perform the backpropagation process for a given neural network architecture.

## 3 Neural-network for ABLT detection

Once trained, the neural network can be treated as a blackbox that receives a portion of a lidar signal $B(z)$ as input (e.g., the first 2000 points of the signal exhibited a good performance in our experiments), and ouputs an approximate ABLT vector index NN $(B(z))$ value such that the ABLT altitude $z_{\text{ABL}}^{\text{NN}}$ is given by:

$$z_{\text{ABL}}^{\text{NN}} = z_{lidar}^{res} \times \text{NN}(B(z)), \quad (9)$$

where $z_{lidar}^{res}$ denotes the vertical spatial resolution of the LiDAR setup.

The ABLT detection was approached as a regression problem, as the NN outputs a single, continuous value on its final layer. A classification approach was also attempted, where the output layer of the NN consisted on a 2000 unit softmax layer, with each element indicating an ABLT occurrence probability at its corresponding altitude. Nevertheless, performance of this second approach was significantly poorer than the direct, regression treatment.

## 4 Aerosol transport during 2019 summer Amazon rainforest wildfires

During August 2019, one of the largest forest fires of recent times was recorded in the Brazilian Amazon forest. The transport of aerosols due to the burning of biomass was registered by satellite information, e.g., from Aqua[13]. This anomalous increase in the aerosol layer reached southwestern Colombia, and was detected at the LiDAR-CIBioFi on-site station, recording the most consistent behaviour on August 21, as reported in figure 2**c** of main paper. Data from NASA's Atmospheric Infrared Sounder (AIRS) instrument, aboard the Aqua satellite[13], as shown in figure 1 of supplementary material (SM), serves as a guide to examine our results. Figure 1-SM shows the transport of carbon monoxide during **a** 14, **b** 15, **c** 16, and **d** 21 of August, at a height of around 5.5 km being dragged towards the northwestern South American. Carbon monoxide is a pollutant that contributes to both air pollution and climate change; the concentrations depicted in colour in figure 1-SM range between green (100 parts per billion by volume-ppbv) and dark red (160 ppbv)[13]. In these plots, we can indeed appreciate the massive aerosol cloud distribution over a vast region of subcontinental South America, and, in particular, the impact on southwestern Colombia, the region where the measurements here reported were taken.

The pollutant transport dynamics portrayed in figure 1 of supplementary material has been contrasted by means of the HYSPLIT model in order to account for wind transport profiles during five days previous to each of the dates plotted in figure 1-SM. The results are shown in figure 2-SM, for **a** August 14, **b** August 15, and **c** August 16, 2019; these dates have been chosen since they allow a further consideration of the results reported in figure 1 and figures 2**a** and 2**b** of main paper, respectively. From these, we conclude that the results of main paper for August 14, 15, and 16 are not directly linked to aerosols arising from the mentioned Amazon forest wildfires, as the wind transport profiles of figure 2-SM indicate that, during these days, the movement of winds mainly comes from regions far away from the Brazilian northwest Amazon region. Indeed, figures 2**a** and 2**b**-SM show winds and associated backward trajectories from the Amazon, but on days when the fire was just beginning in such regions. On the other hand, figure 2**c**-SM shows winds arising from regions completely outside the Brazilian fire area.

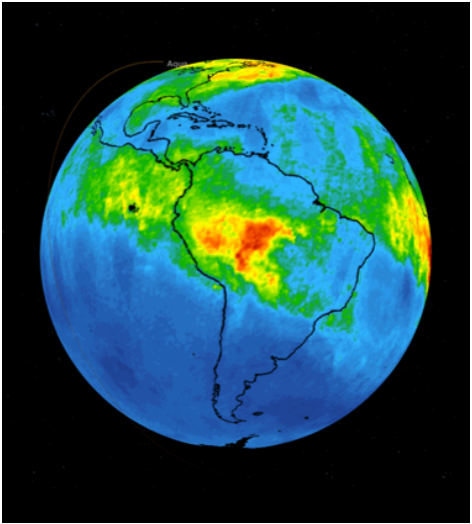 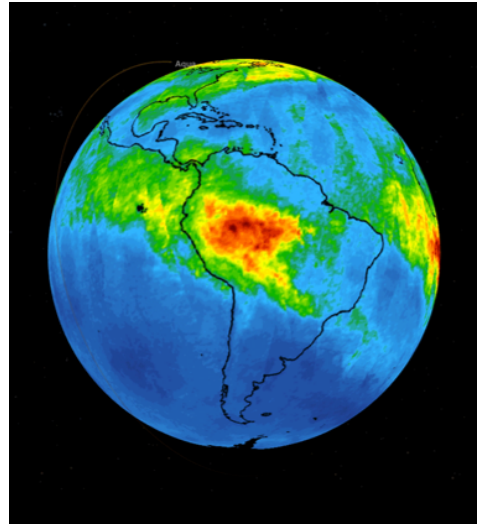

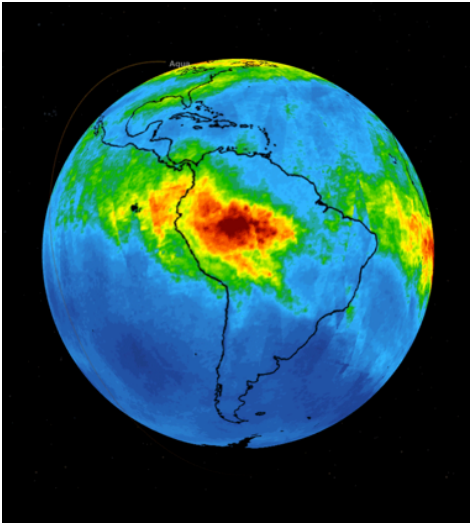 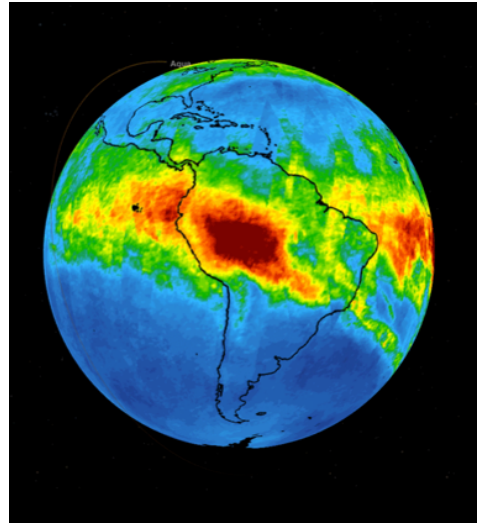

Figure 1: **Aerosol transport during the Amazon rainforest fires, August 2019**. **a**, August 14, **b**, August 15, **c**, August 16, **d**, August 21. Adapted from NASA's Atmospheric Infrared Sounder (AIRS) instrument data, aboard the Aqua satellite. The colours indicate concentrations of carbon monoxide, as follows: Green, at approximately 100 ppbv, yellow at approximately 120 ppbv, and dark red at approximately 160 ppbv[13].

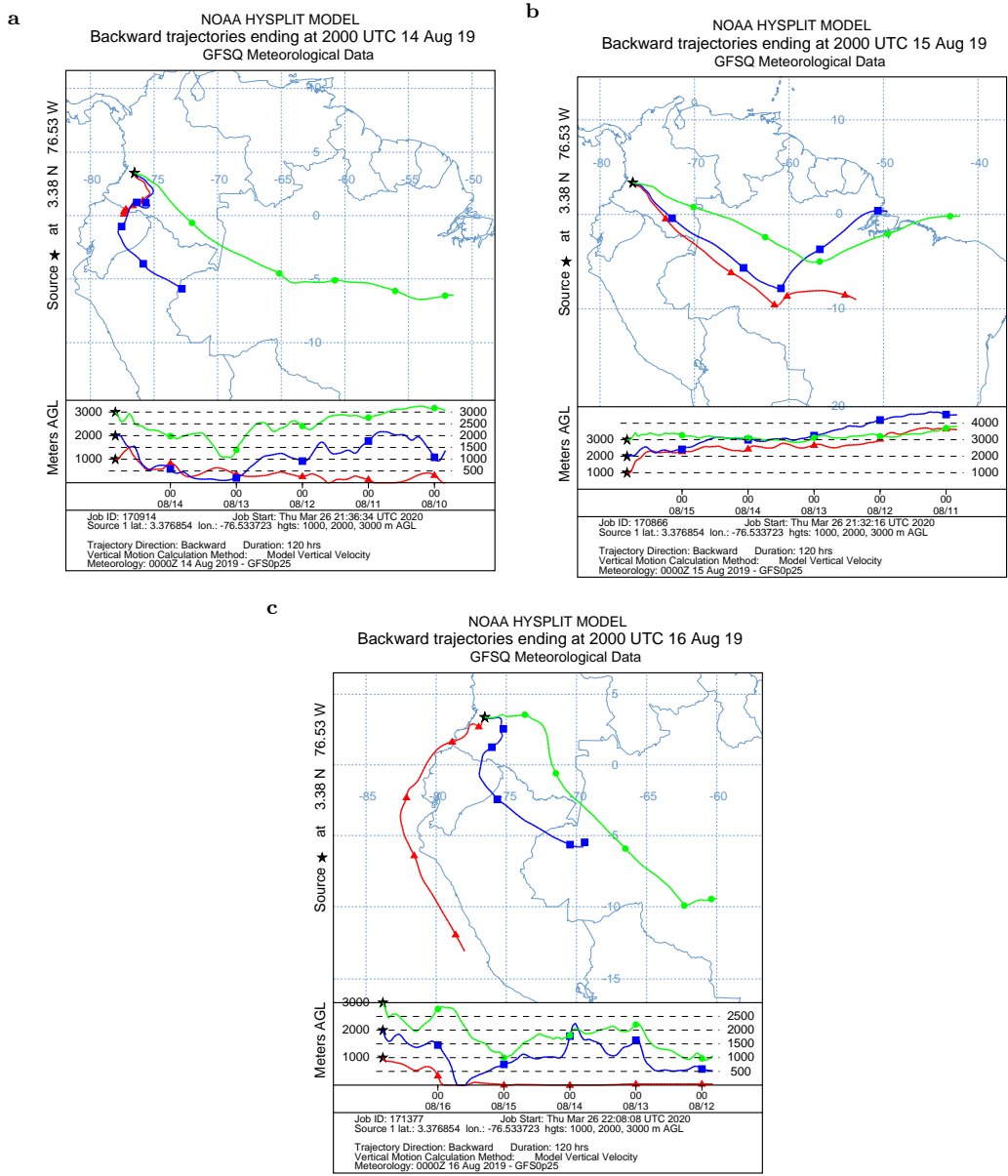

Figure 2: **NOAA HYSPLIT backward trajectories run in ensemble mode.** Five days records and trajectories finishing at 1 km, 2 km and 3 km AGL at 20:00 UTC, on **a**, 14/08/2019, **b**, 15/08/2019, **c**, 16/08/2019. The simulations show the wind transport arising from different regions, to the place marked with a star, where in-situ measurements were taken at the LiDAR station in Cali (Colombia).



\end{document}